\documentclass[prb,aps,preprint,showpacs]{revtex4}
\usepackage{graphicx}% Include figure files
\usepackage{textcomp}% for weird symbols
\begin{document}

\title{Connection Between Magnetism and Structure in Fe Double Chains on the 
Ir$(100)$ Surface}

\author{Riccardo Mazzarello$^{1,2}$}\email[Corresponding author. Present Address: Computational Science, Department of Chemistry and Applied Biosciences,
ETH Zurich, USI Campus, via Giuseppe Buffi 13, CH-6900 Lugano,
Switzerland. Email address: ]{riccardo.mazzarello@phys.chem.ethz.ch} 
\author{Erio Tosatti$^{1,2,3}$}

\affiliation{
$^1$ SISSA, Via Beirut 2/4, 34014 Trieste, Italy \\
$^2$ DEMOCRITOS-INFM, Via Beirut 2/4,~34014 Trieste, Italy \\
$^3$ ICTP, Strada Costiera 11,~34014 Trieste, Italy\\}

\date{\today}

\begin{abstract}
The magnetic ground state of nanosized systems such as Fe double chains, 
recently shown to form in the early stages of Fe deposition on 
Ir$(100)$, is generally nontrivial. Using \emph{ab initio} density functional 
theory we find that the straight ferromagnetic (FM) state typical of bulk Fe 
as well as 
of isolated Fe chains and double chains is disfavored 
after deposition on Ir$(100)$
for all the experimentally 
relevant double chain structures considered. 
So long as spin-orbit coupling (SOC) is neglected, the double chain lowest 
energy state is generally antiferromagnetic (AFM), a state which appears to 
prevail over the FM state due to Fe-Ir hybridization. 
Successive inclusion of SOC 
adds two further elements, namely a magnetocrystalline anisotropy, and a 
Dzyaloshinskii-Moriya (DM) spin-spin interaction, the former stabilizing the 
collinear AFM state, the second favoring a long-period spin modulation. 
We find that anisotropy is most important when the double chain is adsorbed on 
the partially deconstructed Ir$(100)$ -- a state which we find to be 
substantially lower in energy than any reconstructed structure -- so that in 
this case the Fe double chain should remain collinear AFM. 
Alternatively,
when the same Fe double chain is adsorbed 
in a metastable state onto the $(5 \times 1)$ fully reconstructed Ir$(100)$ 
surface, the FM-AFM energy difference is very much reduced and the DM 
interaction is expected 
to prevail, probably yielding a helical spin structure. 
\end{abstract}
\maketitle

\vfill
\eject

\section{Introduction}

Controlling the magnetic order of materials is a long 
standing goal of applied solid state physics, with a tremendous impact on 
the information technology industry. 
The onset of 
a magnetic moment in a transition metal atom
arises primarily out of 
intra-atomic Hund's rules, which are poorly structure-dependent 
even in a solid. 
Interatomic magnetic order, however, depends very critically on structure. 
As is well known, for example, bcc Fe is a prototypical ferromagnetic metal, 
but the magnetic properties do change with the crystal structure and 
the Fe-Fe interatomic distance, so that bulk Fe can support AFM configurations 
in the metastable fcc structure.\cite{abrahams,gonser,tsunoda,li,qian}  
Low-dimensional and mesoscopic systems offer new possibilities to control the 
magnetic order of Fe. In particular, the heteroepitaxial growth of Fe films 
and nanowires on non-magnetic transition metal substrates is expected to 
yield novel magnetic structures due to the combined effects of 
a) lattice mismatch, b) reduced dimensionality, c) hybridization of 
Fe $d$-orbitals with the substrate, and d) spin-orbit related effects 
for heavy metal substrates.

Novel experimental techniques have been developed, such as spin-polarized 
scanning tunneling microscopy (SP-STM),\cite{wiese1} capable of resolving the 
magnetic structure of 
nanosized 
systems at the atomic level. 
This technique has recently shown that the ground state of a single 
monolayer (ML) of Fe on W$(001)$ is a collinear AFM 
state rather than a FM one.\cite{wiese2} One way to partly rationalize the 
demise of ferromagnetism in this system could be the observation that the 
density of states (DOS) at the Fermi energy $n(E_{F})$ is strongly depressed 
upon adsorption.\cite{wiese2} 
Since the FM susceptibility is essentially proportional to  $n(E_{F})$,
while the AFM susceptibility is not, antiferromagnetism might 
happen to suffer less from interaction with the substrate, and prevail 
over ferromagnetism because of that.
This hypothetical possibility fits the additional 
experimental observation that Fe monolayers remain FM on W$(110)$, where 
adsorption is weaker, this different tungsten face being better packed and less
reactive than W$(001)$.\cite{elmers} Single MLs of Fe on Ir$(111)$ have also 
been shown to be AFM and to form complex, collinear mosaic 
structures.\cite{vonbergmann}

Here we are concerned with deposited Fe nanostructures rather than monolayers.
The initial steps of Fe deposition on the $(1 \times 5)$ reconstructed 
Ir$(100)$ surface of Ref.\onlinecite{hammer} showed that Fe deposition 
initially forms metastable double chains, which appear to occupy the 
trough-like double minima of the quasi-hexagonal Ir$(100)$ top layer height 
profile. While the presence 
of the $(1 \times 5)$ periodicity suggests
the permanence of reconstruction or
at least some amount of reconstruction, it does not actually certify 
that the pristine quasi hexagonal reconstruction of the Ir$(100)$ substrate
remains unaltered upon Fe adsorption. 
The existing data do not permit to resolve
the detailed structure of the underlying Ir substrate.\cite{hammer} 
The Fe double chains might deposit without altering the 
initial reconstruction, or they may alter it to some extent. Indeed, it is 
found that 
the $(1 \times 5)$ Ir(100) reconstruction becomes eventually lifted at high Fe 
coverage and high temperature.\cite{hammer}

The scope of the present 
calculations is to analyze and possibly predict the magnetic state of 
Fe double chains adsorbed on Ir$(100)$. 
As an added bonus, we wish to establish whether something can be learnt from 
the relationship between magnetism and structure, 
also 
in view of the 
ongoing SP-STM experiments on these systems at low-temperatures.\cite{wiese3}

This is not the first theory work on Fe double chains on Ir(100). 
In Ref.\onlinecite{spisak1} the structure and energetics of this system was 
already investigated by first-principles density-functional theory (DFT). 
Different adsorption sites were considered and structural relaxations 
were performed for both FM and non-magnetic (NM) configurations. %and the 
The
FM configuration was shown to be always preferred over the NM one,
which is consistent with Fe's strong Hund's rule intra-atomic parameters. 
However, these studies did not examine other interesting possibilities such as 
AFM or non-collinear orderings. Furthermore the effects of SOC were not 
considered, so that no specific easy magnetization axis 
and magnetocrystalline anisotropy parameters
were established.  

We will present here two sets of calculations. The first set will investigate
collinear spin structures only and, for that purpose, we will use a realistic 
model of the substrate consisting of a seven Ir layer slab. 
The ground state energy and optimal state of magnetization of free standing 
and Ir(100) deposited Fe double chains will be compared without SOC, i.e., 
within the scalar relativistic approximation. 
Here only two possible magnetization states are considered,
namely FM and AFM (same magnetization sign of two Fe atoms across the
double chain, alternating sign between first neighbors parallel to the 
chains).
Crucially, different structures will be considered for the underlying Ir(100)
substrate, and their relative energetics compared. 
In a second set of calculations, the SOC will be 
included by switching to the more time-consuming fully relativistic 
approximation, and here different AFM spin directions will be considered, 
so as to extract magnetic anisotropy energies (MAEs). 
For MAE calculations the same realistic seven Ir layer slab will be used to
model the surface. Noncollinear spin structures with opposite chirality
will also be considered, so as to extract the DM coupling energy.
However, because of the larger supercells required along the chain 
direction to model spin spirals, this set of calculations is limited by 
computer time economy to smaller and simpler slabs. 
Eventually, a fairly complete scenario of the structures, energies, and 
magnetization geometries will emerge, allowing a discussion, and a tentative 
prediction subject to our rather limited accuracy, of the relationship between
them. Our tentative conclusion is that Fe double chains metastably 
deposited on fully reconstructed Ir(100) may develop long-pitched helical spin
structures whereas the same double chains on the partly reconstructed surface, 
a state of
much lower energy, should exhibit a simple, collinear AFM ground state.
  
\section{Computational Methods}

Standard DFT electronic structure calculations were carried out within a
GGA approximation with a PBE exchange-correlation functional,~\cite{pbe} 
as implemented in the plane-wave PWscf code included in the 
Quantum-Espresso package.~\cite{pwscf, note} 
We employed ultrasoft pseudo-potentials generated 
according to the Rappe-Rabe-Kaxiras-Joannopoulos scheme.\cite{rrkj} 
The wavefunctions were expanded in plane waves with a kinetic energy cutoff
of 30 Ry, whereas the charge density cutoff was 300 Ry for slab calculations
and 800 Ry for free-standing wires. In all the structural optimization runs,
Hellmann-Feynman forces were 
calculated with high accuracy (at each step, the allowed error in the total 
energy was set to 10$^{-7}$ Ry) and a stringent convergence criterion was used 
for structural energy minimization (all components of all forces required 
to be smaller than 10$^{-3}$ atomic units and the change in the total energy 
between two consecutive steps required to be less than 10$^{-4}$ atomic units). 
Convergence with respect to k-points, smearing parameters,  wavefunction and 
density cutoff was checked very carefully. Furthermore, the total energies and 
forces of the optimized structures were recalculated within the 
projector-augmented wave (PAW) method (same method used in Ref.~\onlinecite{spisak1}), 
very recently implemented in the PWscf code, and found to be in good agreement with 
the ultrasoft pseudopotential calculations. 
Free-standing single and double Fe wires 
in the initial test calculations 
were modeled by chains 
parallel to the $\hat{z}$-axis and periodically repeated along the $x$ and
$y$ direction. The minimum distance between periodic images was 20 a.u. 
For single chains, the intra-chain spacing was allowed to vary so as to 
determine the equilibrium spacing. For double chains, the intra-chain 
spacing was set at 2.758 \AA{}, corresponding to the substrate-imposed 
intra-chain spacing of deposited chains which we will need to adopt in
later calculations.

The reconstructed Ir$(100)$ surface has $(1 \times 5)$ periodicity and 
a $\sim 20\%$ higher lateral density (in its quasi-hexagonal top layer) than
a regular bulk $(100)$ plane with its square lattice. A $(1 \times 5)$ 
supercell was used for the clean Ir$(100)$ and for FM Fe double chains 
on Ir$(100)$, whereas a $(2 \times 5)$ supercell was required 
for the AFM case. In the following, the $\hat{y}$-axis will be taken 
perpendicular to the surface, the $\hat{z}$-axis parallel to the chains,
and the $\hat{x}$-axis in the plane and normal to the chains.
In all scalar relativistic calculations, and in the SO calculations of magnetic
anisotropy, the Ir substrate was modelled as a seven layer slab 
periodically repeated across 13 \AA{} wide vacuum regions. Fe double chains 
were deposited on one face of the slab, while the other face was a perfect 
$(100)$ surface. Both the reconstruction of the clean surface and the 
relaxation of Fe/Ir$(100)$ systems were treated
by allowing the four 
topmost Ir layers to relax, and starting from a six-atom/cell Ir topmost layer.
A few scalar-relativistic 
test 
calculations were repeated with a 9-layer, symmetric 
slab with Fe double chains on both faces of the slab. The agreement between 
these calculations and those carried out with asymmetric 
7-layer
slabs was very good. 
A $2 \times 10 \times 1$ Monkhorst-Pack mesh~\cite{pack} of special points was 
used for the integration over the Brillouin Zone for the $(1 \times 5)$ cell 
and an equivalent mesh was used for the $(2 \times 5)$ cell. The Fermi function
smearing approach of Ref.~\onlinecite{paxton} was used to deal with electron 
occupancy near the Fermi level, with a smearing parameter of $0.01$ Ry. 
As a test of our pseudopotential, we calculated the lattice parameter, 
$a_{0}$, and the bulk modulus $B$ of bulk fcc Ir. 
Our results,
$a_{0}=3.90$ \AA{} and $B=3.42$ Mbar, compare very well with the 
experimental values, $a_{0}=3.84$ \AA{} and $B=3.55$ Mbar.

\section{Results: Structure}

\begin{figure}%[htp]
\begin{center}
\includegraphics[angle=0,width=11.0cm]{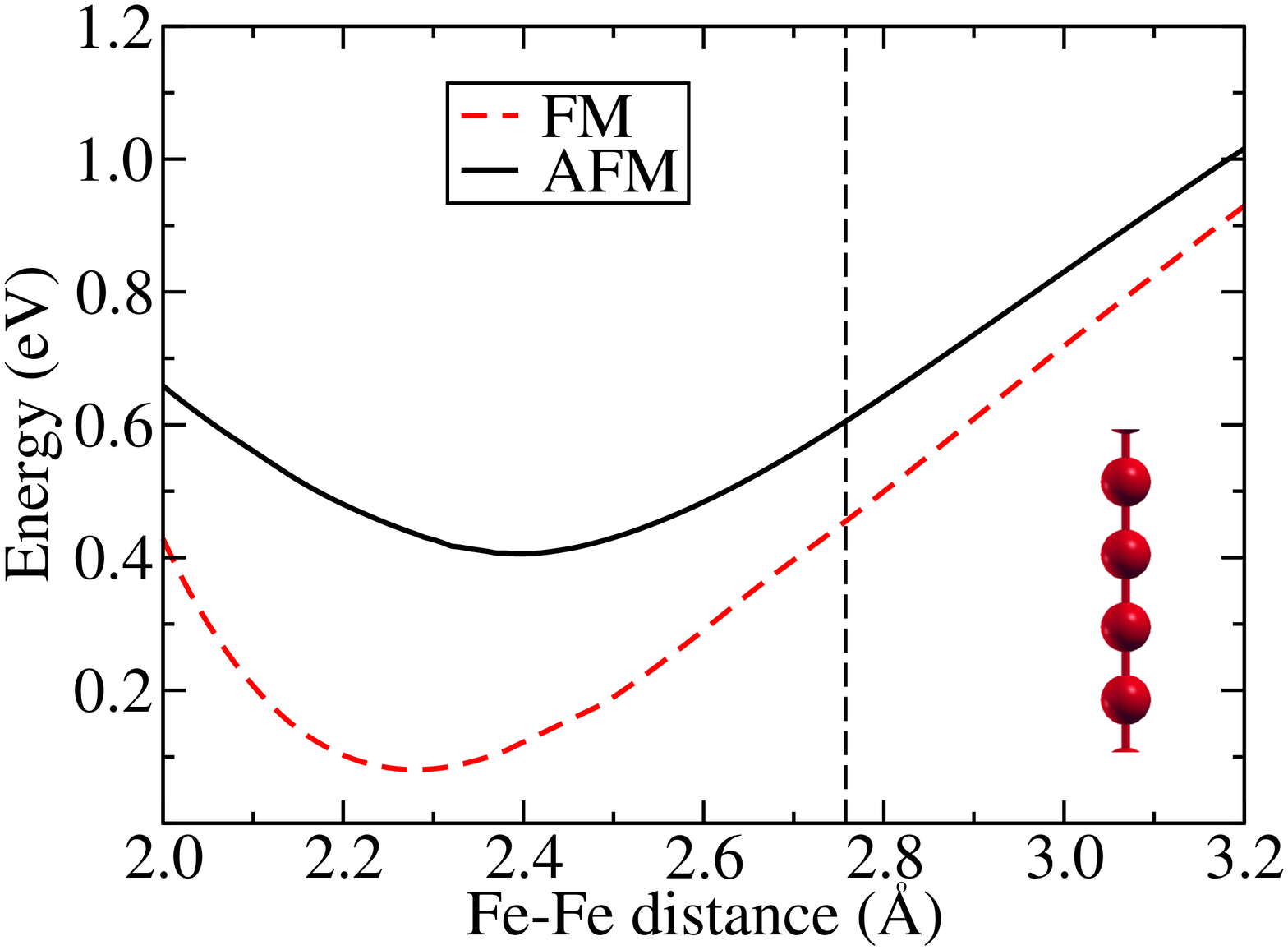}
\caption{Total energy per atom of FM and AFM free standing Fe single 
chains as a function of Fe-Fe distance.The dashed vertical line corresponds 
to the theoretical interatomic distance of the Fe chain 
deposited on Ir$(100)$.}
\label{en_wires_fm_af}
\end{center}
\end{figure}

We started off with ideal, free-standing Fe single chains. Similar to 
previous work,~\cite{spisak2,nautiyal,tung} we found first of all that 
the lowest energy state of free standing Fe chains is
FM. In Table~\ref{tab_eqfes}
the calculated equilibrium Fe-Fe distance and the magnetization per Fe atom 
of NM, FM and AFM chains are compared with those 
given 
in the literature. 
The calculated total energy of free single chains in the 
FM and AFM configuration 
as a function of Fe-Fe distance is shown in Fig.\ref{en_wires_fm_af}.
The epitaxial Fe chains on Ir$(100)$ are stretched lengthwise and the 
energy difference between 
AFM and FM chains, $\Delta E = E_{AFM} - E_{FM}$, shrinks for 
stretched wires. At the theoretical intrachain Fe-Fe distance of wires 
deposited on Ir$(100)$, 2.758 \AA{}, $\Delta E=0.142$ eV/Fe atom. 
 
For double chains, we 
restricted ourselves to the  2.758 \AA{} 
intrachain Fe-Fe 
distance only. It was found that free 
standing Fe double chains are also FM, although 
the energy difference per Fe atom between FM and AFM 
is smaller than for isolated chains. 
Energies and magnetizations of Fe atoms are shown in Table~\ref{tab_fed}. 
The magnetization of an atom is conventionally calculated 
by integrating the up-down spin density difference in a sphere centered on 
the atom, with radius equal to half the distance between the atom and its 
nearest neighbor.

Separately we considered the clean, reconstructed $(1 \times 5)$ Ir$(100)$ 
surface. The calculated surface energy $E_{s}$ and work function $W$ of the 
surface are $1.31$ eV and $5.51$ eV respectively, in excellent agreement with 
experiments~\cite{deboer,rhodin} and
with 
previous theoretical work.~\cite{ge,spisak1} 
The calculated surface energy difference between the perfect $(1 \times 1)$ and
the reconstructed $(1 \times 5)$ surface was $0.05$ eV / ($(1 \times 1)$ area),
which also compares well with previous GGA calculations.~\cite{ge} 
Note that, had we used the LDA approximation, the $(1 \times 5)$ reconstructed
phase would instead have been unstable,~\cite{ge} in contrast to experiments.

The structural parameters of the reconstructed Ir$(100)$ surface are 
shown in Table~\ref{tab_ir} (the notation of Ref.~\onlinecite{spisak1} is 
used). Our results are in good agreement with experimental structure
parameters as measured by LEED,~\cite{leed} as well as with previous 
calculations.~\cite{ge,spisak1}

All basic ingredients ready, we proceeded to investigate the properties and 
energetics of Fe double chains deposited on the Ir substrate. There are three 
different energy scales at play in this system. The first is the structural 
scale, involving energy differences of the order of 100 meV/Fe atom. 
The second is the magnetic intersite exchange scale, involving differences of 
the order of 10 meV/Fe atom. 
The third is the spin orientational scale (spin orbit related),
involving differences of the order of 1 meV/Fe atom.
We stress that the size of inter-site exchange interactions between magnetic 
Fe atoms is two orders of magnitude smaller than the intra-atomic ``magnetic'' 
exchange energy scale, of order of 1 eV/Fe atom, due to the very strong 
Hund's rule intra-atomic interactions.

\begin{figure}%[htp]
\begin{center}
\includegraphics[angle=0,width=11.0cm]{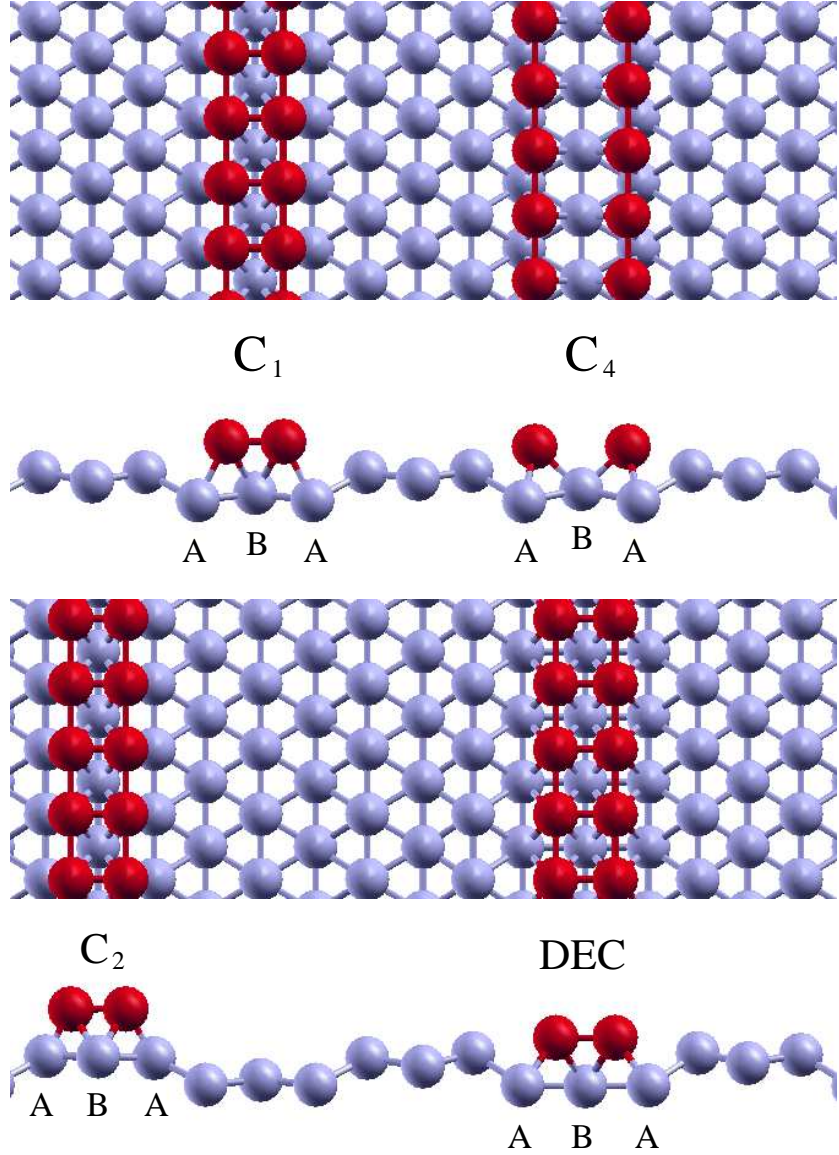}
\caption{Top and side view of the reconstructed Ir$(100)$ surface with the 
studied configurations for the dimer chain. Configurations C$_{1}$  and C$_{4}$
correspond to Fe chains adsorbed on the troughs of $(1 \times 5)$ Ir$(100)$, 
whereas  C$_{2}$ corresponds to Fe chains sitting on the hills 
of $(1 \times 5)$ Ir$(100)$. DEC is the partially deconstructed structure,
wherein the Fe atoms sit on the hollow sites of a quasi-square, locally 
deconstructed Ir$(100)$. A and B indicate the Ir atoms nearest neighbors 
of Fe. Vertical displacements have been exaggerated for clarity purposes.}
\label{config}
\end{center}
\end{figure}

We proceeded to examine structures first. Several configurations of Fe double 
wires on Ir$(100)$ were considered, corresponding to different adsorption 
sites for the Fe atoms (see Fig.\ref{config}). Adopting the notation of 
Ref.~\onlinecite{spisak1} we considered 
C$_{1}$, C$_{2}$ and C$_{4}$ 
configurations. Configurations C$_{1}$  and C$_{4}$ correspond to Fe chains 
adsorbed on the troughs of $(1 \times 5)$ Ir$(100)$, 
whereas  C$_{2}$ corresponds to
Fe chains sitting on the hills of $(1 \times 5)$ Ir$(100)$. 
The zig-zag shaped configuration denoted as C$_{3}$ in 
Ref.~\onlinecite{spisak1} was not considered, for STM images fail to suggest 
zig-zag shaped chains.~\cite{hammer}  

We found, 
interestingly, 
that  C$_{1}$, C$_{2}$ and C$_{4}$ configurations 
were all metastable. This is because the Fe chains should
lift the reconstruction of Ir$(100)$, rather than adsorb on the $(1 \times 5)$ 
fully reconstructed 
structure. In fact the calculated adsorbtion energy of double chains 
on perfect, unreconstructed  $(1 \times 1)$ Ir(100), where the top layer is a 
square lattice, is 0.57 eV/Fe atom larger than on  $(1 \times 5)$ Ir(100), 
where the top layer is a distorted triangular lattice. 
This energy difference was calculated in a grand-canonical definition, 
i.e. by subtracting the sum of the energy of the fully deconstructed structure 
and the energy of a bulk Ir atom to
the energy of the reconstructed structure.
However, the Ir surface deconstruction 
from  $(1 \times 5)$ to  $(1 \times 1)$ implies the removal of ~20 \% of 
the first layer Ir atoms, which may 
not readily take place when the double chains are experimentally 
deposited at low temperature and low coverage.~\cite{hammer}
At high T and/or high Fe coverages, full experimental deconstruction of 
Ir(100) takes place, with expulsion of the excess Ir atoms from the first layer
and formation of Ir chains on top of the surface. 
Similar superstructures consisting
of Ir rows on Ir$(100)$ have recently been seen also in case of
adsorption of H atoms on this surface, 
at sufficiently high temperatures.~\cite{hammer2} 
However, at low T and coverage, the  $(1 \times 5)$ structure may 
be kinetically frozen, 
given the massive atomic migration and rearrangement required 
to produce the $(1 \times 1)$. A second possibility is a partial 
deconstruction of the surface, taking place without removal of any Ir atoms. 
A simple displacement 
of Ir atoms 
from beneath the Fe double chains (where the Ir layer structure may be locally 
altered) to besides the chains should be much less kinetically hindered than 
a full deconstruction. 
To investigate this possibility, we started from C$_{1}$ and C$_{4}$ 
structures and looked for concerted displacements of Fe and Ir atoms that 
would spontaneously lower the energy. The structures were perturbed by 
moving the Fe atoms midway between the C$_{1}$ and C$_{4}$ adsorption sites: 
then they were relaxed using a standard Broyden-Fletcher-Goldfarb-Shanno (BFGS)
quasi-Newton method.
It became apparent in this way that both C$_{1}$ and C$_{4}$ structures are 
unstable against a lateral shifting motion of Ir atoms underneath the 
Fe chains. To lower the energy, the Ir atoms shifted sideways so as to restore 
a square ideal $(100)$ geometry underneath the double chains, and accumulating 
besides them.
This Ir rearrangement yielded a partially deconstructed $(1 \times 5)$ 
structure wherein the Fe atoms sit on the hollow sites of a quasi-square, 
locally deconstructed Ir$(100)$ surface (see Fig.\ref{config}, where the 
structure is denoted as DEC), slightly compressed along the 
direction perpendicular to the Fe chains (the Ir-Ir distance along this 
direction is 2.67 \AA{}, to be compared with the equilibrium distance
of 2.76 \AA{}). The atomic coordinates of this structure 
in the FM phase are listed in Table~\ref{tab_dec}. 
This structure is lowest in energy among the $(1 \times 5)$ Fe/Ir systems
explored, with a large energy gain of about 0.46 eV per Fe atom 
with respect to C$_{4}$, which is the lowest energy
reconstructed structure. 
Since STM images do not yield information on the 
position of Ir atoms beneath or besides the Fe chains, this partially 
deconstructed $(1 \times 5)$ Fe/Ir (DEC) structure seems as compatible as 
reconstructed (REC) C$_{1}$ and C$_{4}$ structures with available data, 
and thus deserves to be investigated on similar grounds. 
We remark 
finally that {\it both} the REC and DEC surface geometries 
are strictly speaking metastable. 
We calculate in fact 
the total energy of double chains 
(coverage 0.4 ML) on a {\it completely deconstructed} Ir$(100)$ surface to be 
still 0.11 eV /Fe atom lower than the energy of the DEC structure, 
and 0.57 eV lower than that of the REC C$_{4}$ structure. 
The larger extent of the latter difference indicates however that 
most of the energy gain is obtained as soon as the Ir rearrangemnent 
is actuated locally beneath the Fe double chain, suggesting that 
structures like DEC should be taken in serious consideration as structural 
candidates. 
The structural parameters for non-magnetic, FM and AFM wires on Ir$(100)$ 
are shown in Table~\ref{tab_feir}. 
We note that, contrary to Ref.~\onlinecite{spisak1}, and surprisingly given the
similarity of approaches, the C$_{2}$ structure is highest
in energy amongst all REC structures, rather than the lowest. We repeatedly
checked all possible sources of error in our calculation but found none. 

\subsection{Ferromagnetism versus Antiferromagnetism}

In agreement with Ref.~\onlinecite{spisak1}, we found for all structures
that nonmagnetic configurations are always disfavored over the magnetic ones, 
reflecting Fe's strong Hund's rule coupling. We then considered in parallel
the REC and DEC structures. In our calculations the lowest energy FM structure 
among the REC ones is C$_{1}$, whereas in Ref.~\onlinecite{spisak1} it 
was  reported to be C$_{2}$, which is least favored in our calculations. 
In the AFM case, 
on the other hand, C$_{4}$ is lower in energy than either C$_{1}$ and C$_{2}$, 
although the structural energy difference between C$_{1}$ and C$_{4}$ 
is quite small, only 0.02 eV. C$_{2}$ is always the highest energy structure, 
which agrees with the experimental evidence that double chains appear to sit 
in the troughs of the $(1 \times 5)$ Ir$(100)$.~\cite{hammer}  However, the 
structural interchain distances of the C$_{1}$, C$_{4}$ and DEC structures 
are 2.42, 4.17 and 2.52 \AA{} respectively, all different from the apparent 
distances in the STM pictures, 3.3 $\pm$ 0.2 \AA{}. 
As %already 
pointed out in Ref.~\onlinecite{spisak1}, the error of this 
``experimental'' value may well be much larger than 0.2 \AA{}
because apparent maxima in STM images may strongly deviate from
actual centers of the Fe atoms. In conclusion, the STM pictures do not
really discriminate between various structures. The calculated magnetic 
moments of Fe atoms are of the order of 3.1 +/- 0.1 $\mu_{B}$ in both 
FM and AFM configurations. The Ir atoms moments neighbouring the magnetic Fe 
chains generally become magnetically polarized, with moments of order 
0.1-0.3 $\mu_{B}$ in the FM case; in the AFM case, on the other hand, some 
of the moments of nearby Ir atoms vanish by symmetry (in the DEC
structure, they {\it all} vanish by symmetry).
 
We also considered AFM structures wherein Fe atoms transverse to
the double chain have opposite magnetization sign (and the coupling
between first neighbors parallel to the chain direction is still AFM).
The energy of this tranverse
AFM configuration is higher than 
the longitudinal AFM
one considered above by $0.04-0.1$ eV per atom
depending on the structure, with the exception of C$_{4}$, 
where the two configurations are practically degenerate. 
This is not unexpected, for the distance 
between chains is large in C$_{4}$ so they weakly interact with each other.
Since these transverse
AFM structures are in general energetically 
higher than the longitudinal AFM ones,
we will not investigate them further.

\begin{figure}%[htp]
\begin{center}
\includegraphics[angle=0,width=8.0cm]{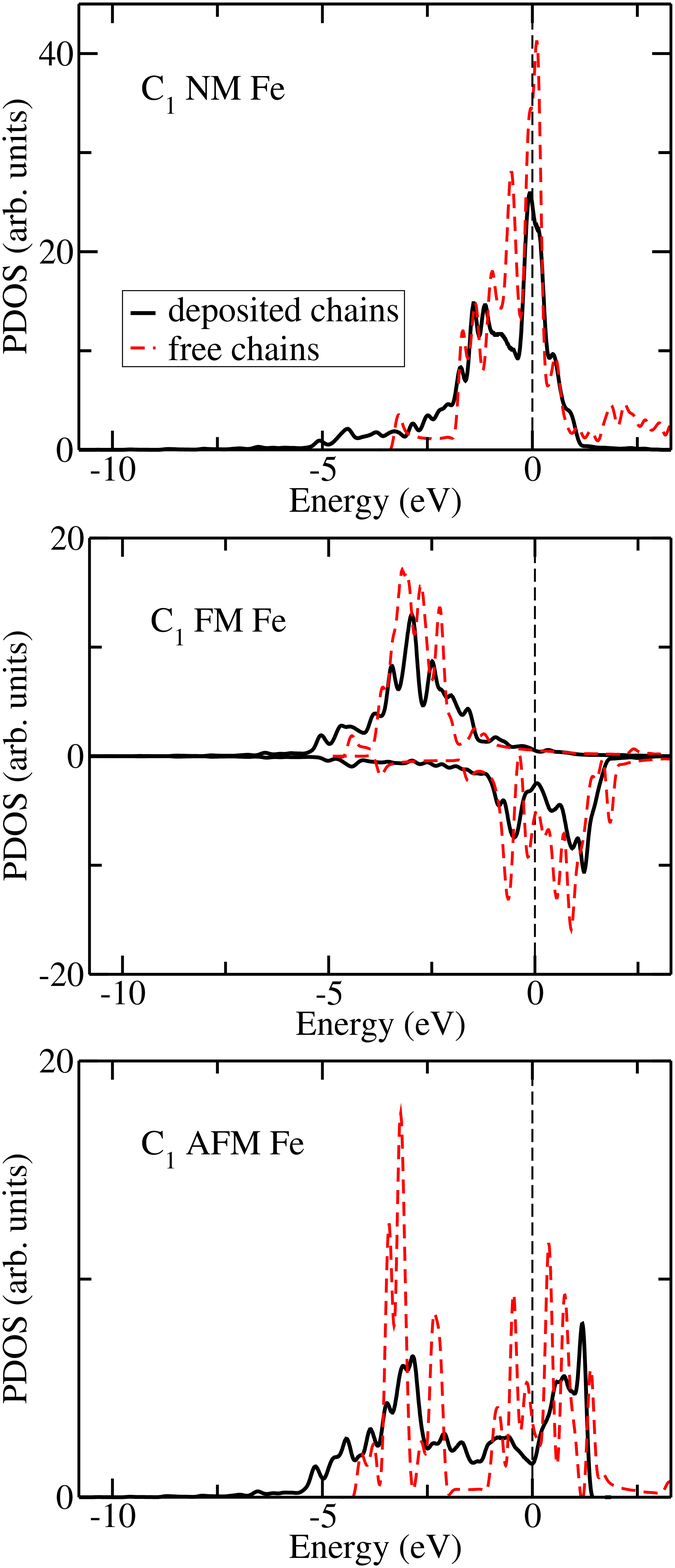}
\caption{Ferromagnetic, antiferromagnetic and non-magnetic density of states 
of the C$_{1}$ REC structure projected onto Fe atoms. 
Dashed lines indicate the DOS of free-standing, double Fe wires.
In the AFM case, the PDOS were calculated by projecting onto all of the 
Fe atoms, i.e. both those with positive magnetic moments and those with
negative ones: as a consequence, the PDOS of spin-up and spin-down electrons 
are the same.}
\label{dosa_c1}
\end{center}
\end{figure}

\begin{figure}%[htp]
\begin{center}
\includegraphics[angle=0,width=8.0cm]{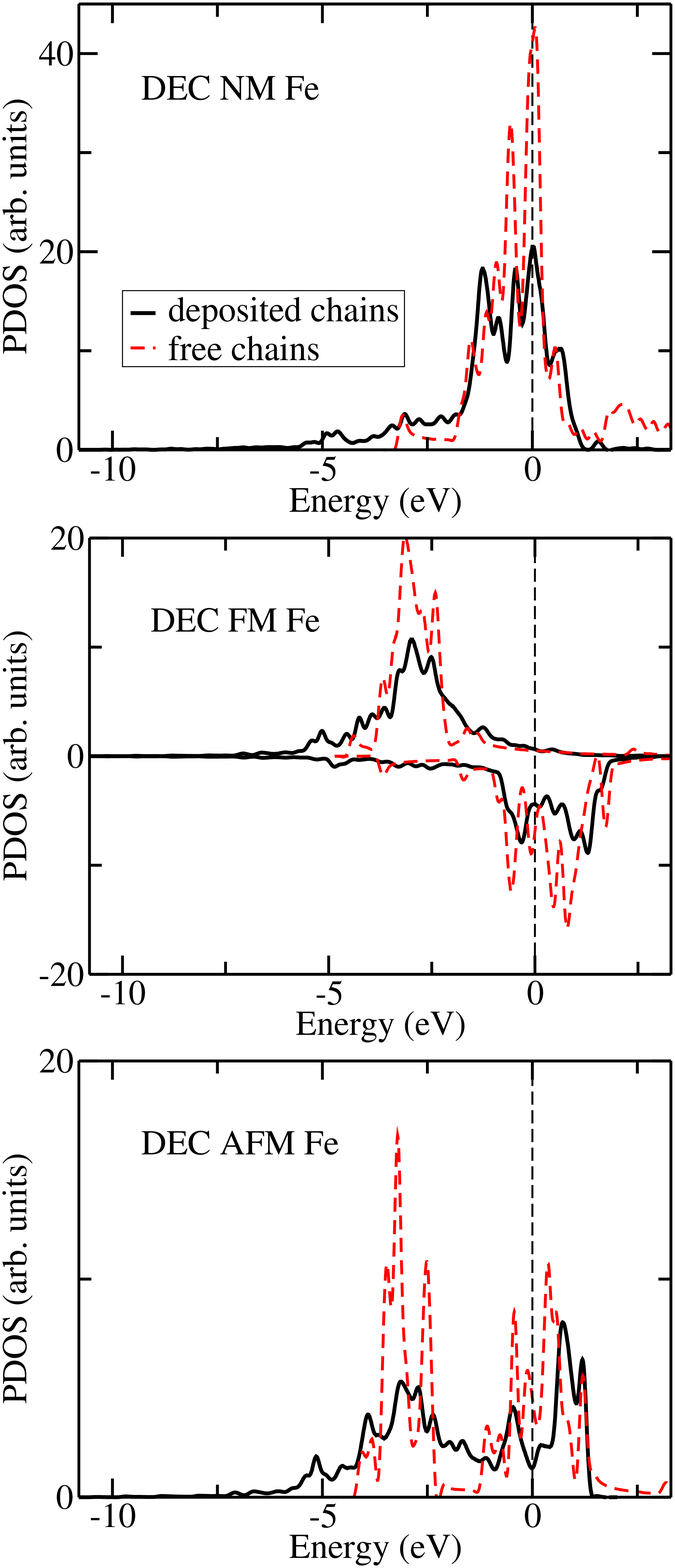}
\caption{Ferromagnetic, antiferromagnetic and non-magnetic 
density of states of the partially deconstructed structure projected onto 
Fe atoms. Dashed lines indicate the DOS of free-standing, double Fe wires.
In the AFM case, the PDOS were calculated by projecting onto all of the 
Fe atoms, i.e. both those with positive magnetic moments and those with
negative ones: as a consequence, the PDOS of spin-up and spin-down electrons 
are the same.}
\label{dosa_dec}
\end{center}
\end{figure}

\begin{figure}%[htp]
\begin{center}
\includegraphics[angle=0,width=11.0cm]{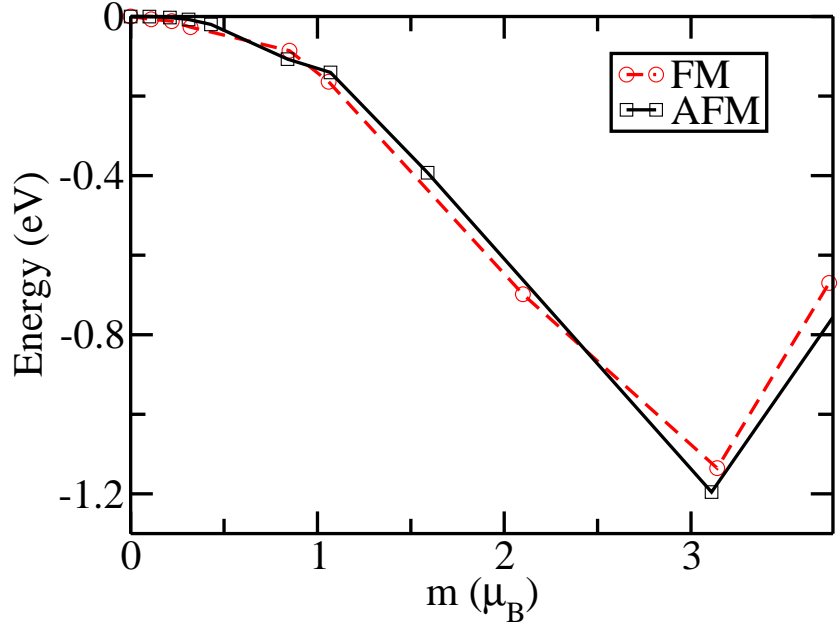}
\caption{Total energy per Fe atom of "toy" DEC FM and AFM structures consisting
of Fe atoms and nearest neighbor Ir atoms as a function
of the magnetic moment $m$ on a Fe atom. These calculations were performed by
adding a penalty functional to the total energy in order to constrain the
local magnetic moment around a Fe atom. For small $m$, which corresponds to
small magnetic fields and small staggered magnetic fields for the FM and AFM
case respectively, the dependence of the energy on $m$ is quadratic and the 
coefficient of the quadratic term is inversely proportional to the FM or AFM
susceptibility.}
\label{suscept}
\end{center}
\end{figure}

The demise of FM in favor of AFM in Ir-deposited chains is due to Fe-Ir 
hybridization, since free-standing double chains are always FM. Following a 
reasoning parallel to that of Bl\"{u}gel et al.\cite{wiese2} this could 
tentatively be rationalized in terms of changes in the respective 
FM and AFM susceptibilities. 
The FM susceptibility should be approximately proportional to the electronic 
density of states (DOS) at the Fermi level evaluated in the NM state and 
projected (PDOS) on the Fe atoms, $n_{Fe}(E_F)$. We calculated the NM DOS for 
all the relevant structures and compared their projected value onto 
Fe atoms with the NM DOS of free-standing, coupled chains 
(see Figs.~\ref{dosa_c1}-~\ref{dosa_dec}, where the NM PDOS of the C$_{1}$ 
and the DEC structures are shown together with the 
FM and AFM ones; the PDOS of the C$_{2}$ and C$_{4}$ structures show a 
qualitatively similar behavior). 
Confirming expectations, we note a clear decrease 
of PDOS upon deposition on the Ir$(100)$ surface.
One might now be tempted to surmise that since $n_{Fe}(E_F)$ is reduced upon 
deposition on Ir$(100)$ due to hybridization with Ir, ferromagnetism might be 
disfavored relative to AFM due to a selective decrease of the FM susceptibility
relative to the AFM one. A PDOS decrease could reduce the violation of 
Stoner's FM criterion $1-Un(E_F)<0$ (where U is an exchange energy parameter), 
while the AFM susceptibility need not do exactly the same, as there is no 
straight a priori proportionality
between PDOS and AFM susceptibility.  
To investigate that aspect, we conducted
constrained magnetization 
calculations allowing a numerical evaluation of the zero-field FM and AFM 
susceptibilities. 
To reduce computational times, we did that for 
a ``toy'' DEC structure consisting of Fe atoms and 
nearest neighbor Ir atoms only 
(in total 2+3 atoms in the FM cell and 4+6 atoms in the AFM cell). 
For this structure the AFM structure is lower in energy than the FM one by
about $60$ meV per Fe atom. The constraint on the Fe local magnetic moments 
(calculated by integrating the magnetization density in a sphere centered on 
the Fe atoms, as explained at the beginning of Section III) was imposed by 
adding a penalty functional to the total energy. 
As Fig.~\ref{suscept} shows, for small magnetizations there is a 
quadratic energy decrease with magnetization, which measures 
separately 
the FM and AFM
susceptibilities. 
The FM and AFM energies remain however extremely close at all small 
magnetizations, 
and do not indicate appreciable differences between FM and AFM
susceptibilities. 
So while there is an Ir-induced FM susceptibility decrease connected 
with the Ir-induced decrease of  $n_{Fe}(E_F)$, that does not seem to explain 
the switch from FM to AFM. 
The AFM energy gain is realized at large magnetization magnitudes,
not revealed at the perturbative level.

The main notable Ir-related difference between FM and AFM states is the finite
magnetic polarization required for the nearby Ir atoms in the FM case,
contrasted by the symmetry-induced zero magnetic polarization for some 
(REC structures) or all (DEC structure) of the nearby Ir atoms in the AFM case.
Due to an interplay between magnetism and structure, the Fe magnetic orbitals 
delocalize over the Ir substrate atoms in the FM case, but less, 
or not at all (depending on the structure), in the AFM case.
As a result the Ir-related reduction of magnetic energy gain is less 
important in the AFM case than in the FM case. 
If this indeed is the mechanism 
that causes the switch from FM to AFM, then it could 
hold for other magnetic 
elements as well. To explore 
this hypothesis, we studied the magnetic properties 
of Mn, Co, Ni double chains on $(1 \times 5)$ Ir$(100)$ 
(restricting to C$_{1}$ and DEC configurations). 
The starting 
unsupported Mn chains were found to be AFM, whereas Co and Ni chains were FM. 
Energy differences between FM and AFM  Mn, Co and Ni double chains 
(free-standing and deposited on Ir$(100)$) are shown in Table~\ref{tab_coni},
with Fe also shown by comparison. 
Similar to the Fe chains, the Ir surface was unstable against deconstruction 
when Mn, Co or Ni chains are adsorbed on the surface. In the end, it turned 
out that Ir-deposited Co and Ni double chains were still FM, 
unlike Fe. However, the 
energy difference between FM and AFM structures was substantially reduced when 
Co and Ni chains were adsorbed on Ir$(100)$ for all geometries 
(except for Co chains in  C$_{1}$ geometry, where it increased by 0.03 eV). 
Double Mn chains remained AFM when 
deposited on Ir$(100)$, but the energy gap between the AFM and FM 
configurations again increased. On the whole, these results seem to confirm our
starting hypothesis. 
We may tentatively conclude therefore that the selective spillout 
of magnetization to Ir atoms near the Fe chains 
present 
in the FM state but reduced or
absent in the AFM state should
play an important role in shifting the energetic balance 
from FM towards AFM, although other, subtler 
and more specific 
effects should  be invoked 
in order to explain the dependence of the relative stability of FM and AFM 
configurations upon the transition metal element and the adsorption structure. 
On this aspect there is room for further work addressing the physical
mechanism in more detail, maybe resorting to some simplified and more 
transparent schemes such as tight-binding.

\subsection{Magnetic Anisotropy}

Magnetic anisotropy energies were calculated for both unsupported and 
deposited Fe double chains, where the REC (C$_{1}$ and C$_{4}$) and
the DEC configurations have been considered. In free AFM Fe double chains, 
the easy axis was found to lie 
along $\hat{y}$, perpendicular to the plane containing the 
chains for chain-chain distances corresponding to the C$_{1}$ and DEC 
structures (see Table~\ref{tab_mae}). 
For large chain-chain distances, the easy axis switched to $\hat{z}$, along the
chains, in agreement with the single chain limit.~\cite{barreteau, mokrousov}
 
In Ir-deposited AFM Fe double chains the easy magnetization axis of both 
C$_{1}$ and C$_{4}$ REC structures was $\hat{x}$, parallel to the surface and 
perpendicular to the chains. In the DEC structure, $\hat{x}$ was instead the 
hard axis, whereas the easy axis was $\hat{z}$, parallel to the chains 
(see Table~\ref{tab_mae}). 
These magnetic anisotropy results hold for FM configurations as well,
as could be expected from phenomenological on-site anisotropy parameters.
From the predicted opposite magnetic anisotropies of REC and DEC structures, 
it follows that the detection of the easy magnetization axis of the double 
chains on Ir$(100)$ by SP-STM techniques could yield indirect but important 
information on the unknown local structure of the Ir$(100)$ surface.

In principle, we note, magnetostatic effects due to magnetic dipolar 
interactions could also give rise to magnetic anisotropy effects. However, 
for our two-chain AFM system these dipole-dipole energies can be estimated 
to be less than 0.1 meV, much smaller than magnetocrystalline energies due 
to SOC, and can be neglected.

\subsection{Dzyaloshinskii-Moriya Interactions}

The second important effect of spin orbit interaction on magnetism is 
the onset of a Dzyaloshinskii-Moriya inter-site interaction term 
of the form\cite{dzyaloshinskii1957,moriya}
\begin{equation}
 H_{DM} = \textbf{D}_{ij} \cdot \textbf{S}_i \times \textbf{S}_j 
\end{equation}
where $\textbf{D}_{ij}$ is the Dzyaloshinskii vector. The DM interaction is 
chiral and is due to the concerted effect of spin-orbit coupling and
a lack of structural inversion symmetry at the surface. The direction of 
$\textbf{D}_{ij}$ is determined solely by structural symmetry.\cite{moriya} 
More specifically, the intrachain inter-site $\textbf{D}_{ij}$ must 
be orthogonal to a mirror plane containing sites $R_{i}$ and $R_{j}$, 
and parallel to a mirror plane bisecting $R_{ij}$. For the Fe double 
chains on Ir(100), where a pair of (magnetically parallel) Fe atoms is 
the effective magnetic site, the vector $\textbf{D}_{ij}$ lies on the surface 
plane and normal to the double chain, i.e. parallel or antiparallel to 
the $\hat{x}$ axis.
The sign of $\textbf{D}_{ij}$, a vector which breaks the left-right structural 
symmetry, will switch by switching the sign of magnetization, 
in accordance with time reversal symmetry. It is otherwise fully determined 
microscopically by the asymmetry of the selfconsistent potential gradient
in the surface region.

We calculated the magnitude and sign of $\textbf{D}$, assumed to be restricted 
to first neighbors, by direct energy difference between 
two noncollinear magnetic 
structures of the deposited double chain, each composed of four Fe pairs, or
eight Fe atom/cell. The magnetization was constrained to be orthogonal between
one Fe pair to the next down the double chain. In the first magnetic structure,
the magnetization direction was taken to rotate in the sense y, z, -y, -z; 
in the second, it counter-rotated in the sense y, -z, -y, z. 
These two magnetic spirals have identical structural, exchange and anisotropy 
energies, so that their energy difference identifies precisely 
the DM term alone.

Since heavy computational cost restricted us to relatively small systems,
we considered two successive sizes, comprising respectively 
12 and 36 Ir atoms, corresponding to Fe nearest neighbor 
atoms and Fe nearest and next nearest neighbor atoms, respectively.
This allowed an appreciation of the kind of finite size error 
involved, as well as some level of extrapolation towards ideally larger sizes.
Atomic relaxations of the small systems were not taken into account, i.e. 
atoms were frozen at the positions obtained by relaxing the corresponding 
7-layer slabs.
Moreover, only two experimentally relevant structures were
considered: the C$_{1}$ (REC) structure and DEC structure.
(As discussed above, the distance betweeen chains in C$_{4}$ is very large 
and somewhat less likely than C$_{1}$).

It turns out that DM favors right-handed cycloidal spin spirals for
both structures.
As far as the REC structure is concerned, the magnitude $D$ of the 
Dzyaloshinskii vector slightly increases for the larger size systems, 
from 2 to 3 meV, whereas anisotropy energies decrease somewhat from
3-4 meV to 1-2 meV.
We conclude that for the REC deposited Fe double chain, 
MAEs are of the order of 1 meV (Table~\ref{tab_mae}),
whereas $D$ is about 3 meV.
In the DEC structure, both $D$ and MAEs 
are large but do
get significantly smaller in the larger size system: $D$ drops from 
12 to 7 meV and $K = K_z-K_y$ from 8 to 2 meV. 
Extrapolating, we conclude that in the DEC deposited double chain
the anisotropy energy $K$ could be about 1 meV, 
$D$ of order 5 meV.

\section{Rotating Magnetism Versus Collinear Antiferromagnetism}

If anisotropy were ideally zero but at the same time the DM term were finite, 
no matter how small, the collinear AFM magnetic structure would spontaneously 
transform to a rotating magnetic structure, whose pitch would diverge as 
$D$ tends to zero.\cite{izyumov} 
On the other hand, once anisotropy is large enough, the collinear AFM state 
will prevail over 
noncollinear magnetism. 
The relatively large 
anisotropies and DM values reported in the previous sections indicate that 
the competition between AFM and helical spin structures needs to be considered 
in quantitative detail, as was recently done for other systems by Bl\"{u}gel
and collaborators.\cite{bode, ferriani, heide}

\begin{figure}%[htp]
\begin{center}
\includegraphics[angle=0,width=11.0cm]{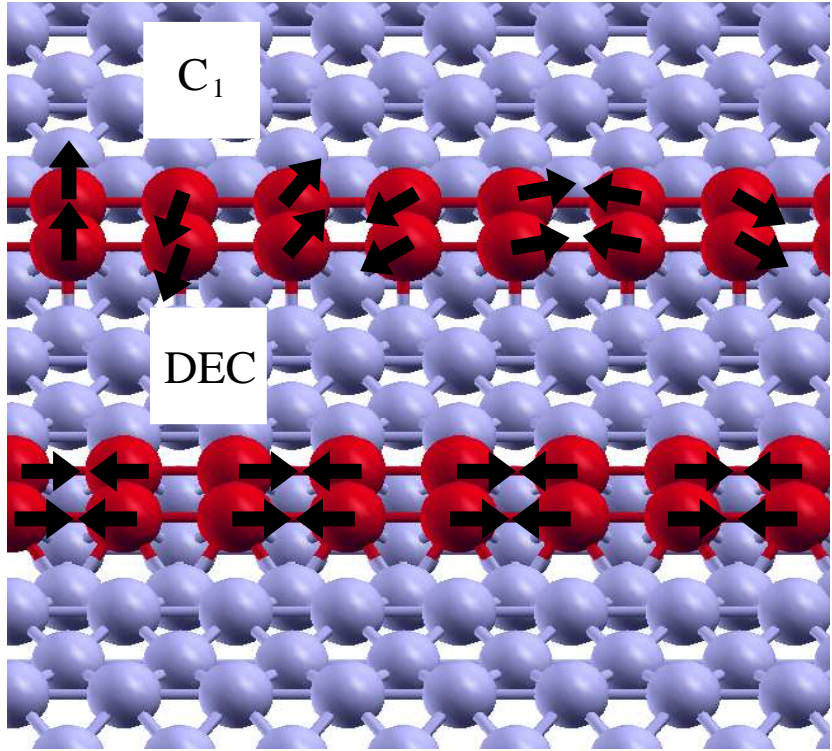}
\caption{View of the spin-structure of the reconstructed C$_{1}$
and the partially deconstructed DEC Ir$(100)$ configuration: Fe magnetic
moments in C$_{1}$ are expected to form a right-handed cycloidal
spin spiral, whereas in DEC a collinear AFM state with moments parallel
to the chains should prevail.}
\label{magn_config}
\end{center}
\end{figure}

In the following we will describe our system by a micromagnetic
continuous model:\cite{brown} for FM systems, this approximation is
justified if the magnetic moment variations are small on a length scale
where exchange and DM interactions are significant. 
For our AFM double chains, given two intrachain nearest neighbor sites 
$i$ and $i+1$ with magnetization $\textbf{m}_{i}$ and $\textbf{m}_{i+1}$, 
a micromagnetic model is valid if
the difference between $\textbf{m}_{i}$ and $-\textbf{m}_{i+1}$ is small.
Within this approximation, taking into account the quasi one-dimensional nature
of our systems, the energy functional is given by
\begin{equation} \label{micro_func}
E= \int_{-\infty}^{+\infty} \left[ A \left( \frac{d\textbf{m}}{dz} \right)^2 + \bar \textbf{D} \cdot \left( \textbf{m} \times  \frac{d\textbf{m}}{dz} \right)+ \textbf{m}^\dag \cdot \bar \textbf{K} \cdot \textbf{m} \right] dz,
\end{equation}
where $A$ is the spin stiffness, $\bar \textbf{D}$ is an effective 
Dzyaloshinskii vector 
and $\bar \textbf{K}$ is an effective anisotropy energy tensor. 
Following the convention usually adopted in micromagnetic calculations,
we assume that $m_x^2+m_y^2+m_z^2=1$.
These three quantities depend on the crystal structure
and can be expressed in terms of the exchange constants $J_{ij}$, 
$\textbf{D}_{ij}$ vectors and 
anisotropy energy tensor $\textbf{K}$ of the discrete model.  
We
may 
assume that only nearest neighbor exchange 
and DM interactions are important: 
then $A \sim d_{intra}J/2$, $\bar \textbf{D} \sim \textbf{D}$
and $\bar \textbf{K} \sim \textbf{K}/ d_{intra}$, 
where $d_{intra}$ is the Fe-Fe intrachain distance
and $J$ and $\textbf{D}$ are the nearest neighbor exchange and DM parameters.
The nearest neighbor $J$ is straightforwardly evaluated from the energy
difference between FM and AFM phases.\cite{noteJ} 
In the following we will separately address two cases, namely:
 
a) $\textbf{D}=D\hat{x}$ parallel to the hard axis. This is the case in the 
DEC surface.

b) $\textbf{D}=D\hat{x}$ parallel to the easy axis. This is the case in the 
REC surface.

a) If $D$ is parallel to the hard axis, then the magnetic moments lie 
in the $(y,z)$ plane, perpendicular to $D$, which contains the double chain
and is orthogonal to the surface. In this case a collinear or two-dimensional
non-collinear structure will appear, depending on the relative strength
of $D$ and the in-plane anisotropy parameter $K \equiv K_z - K_y$, 
where $K_y$ and $K_z$ are the $\hat{y}$ and $\hat{z}$ 
components of the anisotropy energy tensor. This problem is considered in 
detail in the literature\cite{izyumov,dzyaloshinskii1965} and excellently 
summarized in the thesis of M. Heide.\cite{heide_thesis} 
For spin structures lying in the $(y,z)$ plane, 
Formula (\ref{micro_func}) simplifies to
\begin{equation}
E = \int_{-\infty}^{+\infty} \left[ A \left( \frac{d\phi}{dz} \right) ^2 + \bar D \frac{d\phi}{dz} + \bar K \sin^2 \phi \right] dz, 
\end{equation}
where $\phi$ is the angle between the local magnetization and the easy axis, 
$\hat{z}$, and $\bar K \equiv \bar K_{y} - \bar K_{z}$.
A non-collinear, helical state will appear if the DM-related energy gain is 
higher than twice the formation energy of an optimal domain wall in the 
$(y,z)$ plane.\cite{izyumov,dzyaloshinskii1965} This is equivalent to:
\begin{equation}
D > \frac{4}{\pi} \sqrt{\frac{JK}{2}}.
\end{equation}
Inserting the numerical values $J$ = 29 meV and $K$ = 1.7 meV, we obtain 
the inequality $D > 6.3$ meV, estimated for the occurrence of a 
helical state in the DEC structure. From our calculations, we estimate $D$ 
around 5 meV, smaller than the critical 
value, although generally of the same order of magnitude. 
Therefore, we tentatively conclude that the AFM collinear state is 
most likely in the DEC  structure. In view of our poor accuracy, however, 
we cannot totally exclude a helical state with a very long pitch, 
consisting of wide antiferromagnetic domains 
separated by well separated domain walls.

b) if $D$ is paralled to the easy axis, then 
three-dimensional non-collinear structures are also 
possible, besides
collinear and two-dimensional non-collinear ones.
A thorough description
of this case can be found in 
Ref.~\onlinecite{heide_thesis}. 
Since the condition $m_x^2+m_y^2+m_z^2=1$ holds, 
Formula (\ref{micro_func}) can be written as
\begin{equation} \label{micro_func_cart}
E = \int_{-\infty}^{+\infty} \left[ A \left( \frac{d\textbf{m}}{dz} \right)^2 - \bar D \left( m_y \frac{dm_z}{dz} - m_z \frac{dm_y}{dz} \right) + \left( \bar K_{y} - \bar K_{x} \right) m_y^2 + \left( \bar K_{z} - \bar K_{x} \right) m_z^2 \right] dz,
\end{equation}
where $\bar K_{x}$, $\bar K_{z}$ and $\bar K_{y}$ are the easy, intermediate
and hard components of the anisotropy energy tensor respectively 
(see anisotropy energies for the C$_1$ REC structure in Table~\ref{tab_mae}).
Expanding the integrand of (\ref{micro_func_cart}) around the AFM solution, 
$m_y = m_z = 0$, one gets the following Euler-Lagrange equations:
\begin{eqnarray}
A \; \frac{d^2 m_y}{dz^2} + \bar D \frac{dm_z}{dz} - \left( \bar K_{y} - \bar K_{x} \right) m_y & = & 0\\
A \; \frac{d^2 m_z}{dz^2} - \bar D \frac{dm_y }{dz}- \left( \bar K_{z} - \bar K_{x} \right) m_z & = & 0.
\end{eqnarray}
Considering again only nearest neighbor $J$ and $\textbf{D}$, 
these equations have a periodic solution, 
\begin{eqnarray}
m_y & = & \alpha_y \cos (\omega z + \beta_y),\\
m_z & = & \alpha_z \cos (\omega z + \beta_z),
\end{eqnarray}
if and only if~\cite{heide_thesis} 
\begin{equation} \label{Dhard}
D > \sqrt{\frac{J}{2}(K_z-K_x)} + 1.
\end{equation}
Moreover, when the above inequality is fulfilled, the 
non-collinear state is always lower in energy than the AFM collinear solution. 
Therefore, when $D = \sqrt{\frac{J}{2}(K_z-K_x)} + 1$, a second-order phase 
transition to a 3-dimensional state takes place. 
The system undergoes a 
second-order transition to a 2-dimensional helical state in the 
$(y,z)$ plane at slightly higher values of $D$, but this critical point cannot 
be determined analytically.~\cite{heide_thesis} 
We should emphasize that, in our case, the range of $D$ values where 
the 3-dimensional state is stable (which depends on the magnitude 
of the components of the magnetic anisotropy tensor, 
see Ref.~\onlinecite{heide_thesis}) is very narrow. 
Inserting the numerical values $J$ = 0.5 meV and $K_z-K_x$ = 0.7 meV 
corresponding to C$_1$ in Formula (\ref{Dhard}), 
we obtain that the AFM will be destabilized if $D>1.4$ meV. 
Since our REC surface calculations
suggest $D$ values around 3 meV, which is larger than this threshold, 
we conclude that in a REC structure like C$_{1}$, where the double chains do 
not deconstruct the underlying Ir(100) surface, the magnetic ground state
should be non-collinear, and in particular a $(y,z)$ helical state is the
most likely outcome. 
In conclusion, 
a sketch of the predicted magnetic ground state for 
the DEC and REC (C$_{1}$) structures is shown in Fig.~\ref{magn_config}.

\section{Discussion and Conclusions}

We studied by \emph{ab initio} electronic structure and total energy 
calculations 
Fe double chains on  $(1 \times 5)$ Ir$(100)$. Several different structures
with the experimentally observed  $(1 \times 5)$ periodicity were considered,
particularly one, C$_{1}$ REC, where the underlying Ir surface remains 
quasi-hexagonally reconstructed, and another, DEC, where it is partially 
deconstructed, with a large decrease of total energy. 
By addressing magnetism first without spin orbit effects, we find that in all 
structures considered the deposited Fe double chains do not remain FM as in 
vacuum, but generally adopt an AFM ground state. 
The demise of ferromagnetism is attributed to Fe-Ir hybridization. 
The hybridization of Fe with the Ir substrate brings about first of all a 
drop of the Fe-projected density of electronic states near $E_F$ in the 
non-magnetic state, which reduces the FM susceptibility. 
However, we find that the AFM susceptibility is also reduced by the same
amount upon adsorption. 
At large magnetization, AFM appears eventually to be favored by a 
magnetization node intervening by symmetry in the bridging Ir atoms, 
a node which is absent in the FM case.   
By including spin orbit in the calculations, the magnetic anisotropy energies 
of relevant REC and DEC structures have been determined. The easy axis is found
to lie in the surface plane and perpendicular to the Fe double chain in the REC
structure, and parallel to the chains in the DEC structure. Finally,
we calculated the Dzyaloshinskii-Moriya spin-spin interaction energy, and
found it to be generally of a competitive magnitude when compared to 
anisotropy. The different possibilities arising for the resulting ground state 
magnetization pattern are examined. Within the substantial uncertainties 
connected with our estimated computational and finite size errors, we conclude 
that a collinear AFM state with in-plane magnetization vector is likely to 
prevail in the DEC structure, whereas a long period rotating magnetization in 
an orthogonal plane could instead prevail in the REC structure.  
These predictions and clear signatures should be of value for 
future experimental observations by SP-STM techniques.\\ 

\section{Acknowledgments}

We are grateful to R. Wiesendanger for providing the
initial motivation to conduct this study. We warmly acknowledge many
instructive discussions with, and much help by, Andrea Dal Corso. Discussions 
with S. Bl\"{u}gel and R. Gebauer were also stimulating. The work was directly 
sponsored by MIUR PRIN/Cofin 
Contract No. 2006022847 as well as by INFM/CNR ``Iniziativa transversale
calcolo parallelo". The research environment provided by the independent ESF 
project CNR-FANAS-AFRI was also very useful. 
Part of the calculations were carried out on the
SP5 machine at CINECA, Casalecchio.

\newpage

\begin{table}
\begin{tabular}{|c|cc|cc|cc|cc|}
\hline \hline & our work & & Ref.~\onlinecite{spisak2} & & Ref.~\onlinecite{nautiyal} 
& & Ref.~\onlinecite{tung} & \\
\hline
     & $a_0$ (\AA{}) & $m$ ($\mu_{B}$) & $a_0$ (\AA{}) & $m$ ($\mu_{B}$) & $a_0$ (\AA{}) & $m$ ($\mu_{B}$) & $a_0$ (\AA{}) & $m$ ($\mu_{B}$) \\
\hline
FM   & 2.28 & 3.32 & 2.25 & 3.34 & 2.28 & 2.98 & 2.25 & 3.41 \\
AFM  & 2.40 & 3.14 & 2.38 & 3.05 &   -  &  -   & 2.15 & 1.82 \\
NM   & 1.91 & 0.00 & 1.94 & 0.00 &   -  &  -   & 1.94 & 0.0 \\
\hline \hline
\end{tabular}
\caption{ Equilibrium Fe-Fe distance $a_0$ and 
magnetization per atom $m$ for FM and AFM free-standing single wires: 
comparison between our results and recent results.
All results in this Table were obtained by using GGA functionals.}
\label{tab_eqfes} 
\end{table}

\begin{table}[p]
\begin{tabular}{|ccccc|}
\hline \hline $d_{inter}$ (\AA{}) & magnetic struct. & $\Delta E$ (eV) & $E_{int}$ (eV) & $m$ ($\mu_{B}$) \\
\hline
-         &  FM   &  0.000  &   -     &  3.45 \\
(SC)      &  AFM  &  0.142  &   -     &  3.38 \\
          &  NM   &  2.228  &   -     &  0.00 \\
\hline
2.33      &  FM   &  0.000  &  0.835  &  3.14 \\
          &  AFM  &  0.070  &  0.907  &  3.16 \\
          &  NM   &  1.544  &  1.519  &  0.00 \\
\hline
2.40      &  FM   &  0.000  &  0.803  &  3.18 \\ 
          &  AFM  &  0.092  &  0.854  &  3.21 \\
          &  NM   &  1.631  &  1.400  &  0.00 \\
\hline
2.52     &  FM   &  0.000  &  0.733  &  3.25 \\
         &  AFM  &  0.116  &  0.759  &  3.29 \\
         &  NM   &  1.748  &  1.212  &  0.00 \\
\hline
4.14      &  FM   &  0.000  &  0.083  &  3.43 \\
          &  AFM  &  0.090  &  0.136  &  3.37 \\
          &  NM   &  2.211  &  0.101  &  0.00 \\
\hline \hline
\end{tabular}
\caption{ Calculated energy differences per atom, $\Delta E$, 
and magnetizations per atom, $m$, for NM, FM and AFM free-standing
single chains (SC) and NM, FM and AFM double chains at interchain distances 
$d_{inter}$ of $2.33$, $2.40$, $2.52$ and 
$4.14$ \AA{} (corresponding to the interchain distances of Fe double chains 
in the C$_{2}$, C$_{1}$, DEC and C$_{4}$ 
configurations respectively).
The intrachain Fe-Fe distance is 2.758 \AA{} for all the structures.
For each structure, $\Delta E$ is given with respect to the preferred 
FM solution. For double chains the interaction energy between chains, 
$E_{int}$, is shown as well.}
\label{tab_fed} 
\end{table}

\begin{table}
\begin{tabular}{|c|c|c|c|c|}
\hline \hline & Present Work & LEED & Ref.~\onlinecite{spisak1} & Ref.~\onlinecite{ge} \\
\hline
 $d_{0}$        &  1.95  &  1.920 &  1.943 &  1.916 \\
%                &       &        &        &         \\
 $d_{12}$       &  1.96  &  1.94  &  2.00  &  1.97  \\
 $<d_{12}>$     &  2.26  &  2.25  &  2.25  &   -    \\
 $b_{1}^{13}$   &  0.22  &  0.25  &  0.20  &   -    \\
 $b_{1}^{23}$   &  0.54  &  0.55  &  0.47  &  0.47  \\
 $b_{1}^{34}$   & -0.21  & -0.20  & -0.17  & -0.20  \\
 $p_{1}^{2}$    & -0.04  & -0.05  & -0.03  & -0.05  \\
 $p_{1}^{3}$    & -0.07  & -0.07  & -0.07  & -0.03  \\
%                &       &        &        &         \\
 $d_{23}$       &  1.83  &  1.79  &  1.85  &  1.92  \\
 $<d_{23}>$     &  1.91  &  1.88  &  1.89  &   - \\
 $b_{2}^{13}$   &  0.05  &  0.07  &  0.03  &   - \\
 $b_{2}^{23}$   &  0.10  &  0.10  &  0.05  &   - \\
 $p_{2}^{2}$    &  0.03  &  0.01  &  0.00  &   - \\
 $p_{2}^{3}$    &  0.01  &  0.02  &  0.01  &   - \\
%                &       &        &        &      \\
 $d_{34}$       &  1.88  &  1.83  &  1.91  &   - \\
 $<d_{34}>$     &  1.96  &  1.93  &  1.96  &   - \\
 $b_{3}^{13}$   &  0.08  &  0.10  &  0.05  &   - \\
 $b_{3}^{23}$   &  0.04  &  0.05  &  0.02  &   - \\
 $p_{3}^{1}$    & -0.01  &    -   &  0.01  &   - \\
 $p_{3}^{2}$    &  0.00  &    -   &  0.00  &   - \\
%                &       &        &        &      \\
 $d_{45}$       &  1.94  &  1.89  &  1.92  &   - \\
 $<d_{45}>$     &  1.96  &  1.91  &  1.93  &   - \\
 $b_{4}^{13}$   &  0.05  &  0.06  &  0.03  &   - \\
 $b_{4}^{23}$   &  0.02  &  0.03  &  0.01  &   - \\
 $p_{4}^{2}$    &  0.00  &    -   & -0.01  &   - \\
 $p_{4}^{3}$    & -0.01  &    -   & -0.01  &   - \\
\hline \hline
\end{tabular}
\caption{Calculated and experimental structural 
parameters (in \AA{}) of the reconstructed Ir$(100)$ surface.
The notations of Ref.~\onlinecite{spisak1} are used: $d_{0}$ is
the bulk interlayer distance, $d_{ij}$ and $<d_{ij}>$ are the shortest 
and average distance between layer $i$ and $j$, 
$b_{i}^{kl}$ and $p_{i}^{k}$ are vertical and lateral displacement 
amplitudes of atoms $k$ and $l$ in layer $i$.}
\label{tab_ir} 
\end{table}

\begin{table}
\begin{tabular}{|cccc|cccc|}
\hline \hline 
Element & x (\AA{}) & y (\AA{}) & z (\AA{}) & Element & x (\AA{}) & y (\AA{}) & z (\AA{})\\
\hline \hline 
Fe  &     9.537 & 13.424 &  1.365 &  Ir  &     1.379 &  9.686 &  1.374 \\
Fe  &     7.015 & 13.424 &  1.367 &  Ir  &    11.055 &  7.823 & -0.004 \\
Ir  &    13.043 & 12.477 &  1.370 &  Ir  &     8.273 &  7.781 & -0.003 \\
Ir  &     8.274 & 11.664 &  2.747 &  Ir  &     5.492 &  7.823 & -0.004 \\
Ir  &     3.505 & 12.476 &  1.370 &  Ir  &     2.773 &  7.819 & -0.004 \\
Ir  &    10.948 & 11.700 & -0.011 &  Ir  &    -0.015 &  7.819 & -0.004 \\
Ir  &     5.600 & 11.701 & -0.010 &  Ir  &    12.413 &  5.852 &  1.376 \\
Ir  &     1.379 & 11.905 & -0.009 &  Ir  &     9.665 &  5.866 &  1.377 \\
Ir  &    12.424 &  9.780 &  1.371 &  Ir  &     6.881 &  5.866 &  1.377 \\
Ir  &     9.656 &  9.703 &  1.373 &  Ir  &     4.133 &  5.852 &  1.376 \\
Ir  &     6.891 &  9.704 &  1.373 &  Ir  &     1.379 &  5.872 &  1.375 \\
Ir  &     4.124 &  9.780 &  1.372 &  &  &   &  \\
\hline \hline 
\end{tabular}
\caption{Coordinates of the DEC structure in the FM phase: only Fe atoms and Ir
atoms belonging to the four uppermost layers are listed. The $y$-axis is
perpendicular to the surface, whereas the $z$-axis is along the chains.
In the NM and AFM DEC phases, positions of the Ir atoms do not differ
appreciably from those in the FM phase.}
\label{tab_dec} 
\end{table}

\begin{table}

\begin{tabular}{|cccccccc|}
\hline \hline & & $\Delta E$ (eV) & chain-chain d (\AA{}) & Fe- Ir$_{\mathrm{A}}$ d (\AA{}) & Fe- Ir$_{\mathrm{B}}$ d (\AA{}) & $m$ ($\mu_{B}$) & 
$E_{int}$ (eV) \\
\hline
 C$_{1}$ &  NM   &  1.543   &  1.97  &  2.53  &  2.60  &  0.00  &  4.277 \\
         &  FM   &  0.481   &  2.40  &  2.51  &  2.62  &  3.07  &  3.112 \\
         &  AFM  &  0.480   &  2.42  &  2.49  &  2.59  &  3.06  &  3.254 \\
\hline
 C$_{2}$ &  NM   &  1.731   &  1.97  &  2.32  &  2.71  &  0.00  &  4.089 \\
         &  FM   &  0.903   &  2.33  &  2.40  &  2.71  &  3.02  &  2.690 \\
         & AFM   &  0.895   &  2.31  &  2.38  &  2.70  &  2.99  &  2.840 \\
\hline
 C$_{4}$ &  NM   &  1.642   &  3.77  &  2.30  &  2.45  &  0.00  &  4.179 \\
         &  FM   &  0.521   &  4.14  &  2.52  &  2.59  &  3.14  &  3.071 \\
         & AFM   &  0.459   &  4.17  &  2.49  &  2.55  &  3.15  &  3.275 \\
\hline
 DEC     &  NM   &  0.949   &  2.40  &  2.53  &  2.47  &  0.00  &  4.872 \\
         &  FM   &  0.059   &  2.52  &  2.62  &  2.57  &  3.06  &  3.533 \\
         & AFM   &  0.000   &  2.57  &  2.58  &  2.55  &  3.02  &  3.734 \\
\hline
\hline \hline
\end{tabular}
\caption{ Calculated structural parameters and energetics 
for NM, FM and AFM double chains
on the $(1 \times 5)$ Ir$(100)$ surface.
C$_{1}$, C$_{2}$ and  C$_{4}$ configurations, 
where the Ir surface is reconstructed (REC), are considered, as well as
the structure where Fe chains sit on a partially deconstructed surface (DEC).
The energy differences $\Delta E$ are given with respect to AFM DEC, 
which is 
the lowest energy configuration. 
In columns 3 and 4 the distances between Fe atoms and their nearest neighbor 
Ir$_{\mathrm{A}}$ and Ir$_{\mathrm{B}}$ atoms 
(as indicated in Fig.\ref{config}) are provided. 
The interaction energy ($E_{int}$) for a given 
structure and magnetic configuration is defined as the difference between 
the total energy of the structure and the sum of the energies of the 
clean reconstructed $(1 \times 5)$ Ir$(100)$ and twice the energy of the 
isolated Fe chain (with the same magnetic configuration).}
\label{tab_feir} 
\end{table}

\begin{table}
\begin{tabular}{|c|cc|cc|cc|cc|}
\hline \hline & Mn & & Fe & & Co & & Ni & \\
\hline
              & free & dep. & free & dep. & free & dep. & free & dep.\\
\hline
 C$_{1}$ & 0.085 & 0.169 & -0.092 & 0.001 & -0.057 & -0.088 & -0.041 & -0.012 \\
 DEC     & 0.090 & 0.143 & -0.116 & 0.059 & -0.098 & -0.068 & -0.045 & -0.011 \\
\hline
\hline \hline
\end{tabular}
\caption{Energy difference (per adsorbed metal atom) 
between FM and AFM configurations of free-standing and deposited 
Mn, Fe, Co and Ni double wires. 
C$_{1}$ and DEC configurations have been considered.
Energies are in units of eV.}
\label{tab_coni}
\end{table}

\begin{table}
\begin{tabular}{|ccc|c|c|}
\hline \hline & & & E$_{z}$ - E$_{x}$ ($10^{-3}$ eV) & E$_{z}$ - E$_{y}$ ($10^{-3}$ eV) \\
\hline
free FM  & single chain &           &  1.9 &  1.9 \\
free AFM & double chain & ($d_{inter}$ = 2.4 \AA{})      &  0.5 & -0.6 \\
dep. AFM & double chain & (C$_{1}$) & -0.7 &  0.6 \\
dep. AFM & double chain & (C$_{4}$) & -0.8  & 0.7  \\
dep. AFM & double chain & (DEC)     &  1.9 &  1.7 \\
\hline
\hline \hline
\end{tabular}
\caption{Magnetic anisotropy energies (per atom) of 
unsupported FM single chains, unsupported double AFM Fe chains 
at $d_{inter}$ = 2.4 \AA{} and double AFM Fe chains deposited onto 
the Ir$(100)$ surface for REC (C$_{1}$ and C$_{4}$) and DEC configurations.
The intrachain Fe-Fe distance is 2.758 \AA{} for all the structures. 
The $z$-axis is along the chains, whereas the $x$-axis is perpendicular to the 
chains and parallel to the plane containing the chains.}
\label{tab_mae}
\end{table}


\begin{thebibliography}{99}
\bibitem{note} Some of the effects involving small spin-orbit related
energies were obtained and checked independently for accuracy using perturbation theory.
\bibitem{noteJ} It should be noted that the value of J is very small, 
especially in the C1 REC structure.  
The reason why AFM competes with and narrowly wins 
over with FM is discussed earlier in the paper. 
Our tentative conclusion is that it is related to the unfavourable circumstance
that a bridging Ir atom must acquire a spin polarization in the FM state, 
whereas spin polarization has a node at that atom in the AFM state. 
Of course with J values in the meV range there can be no order, even
of the Kosterlitz-Thouless type, unless temperature is hundreds of 
milliKelvin -- which however is where spin-polarized STM experiments 
are usually performed.
\end{thebibliography}
\end{document}